\documentclass[prl,twocolumn,floatfix,amsfonts,superscriptaddress]{revtex4}
\usepackage{graphicx}

\begin{document}

\title{Evidence for non-conventional pairing in Na$_x$CoO$_2$$\cdot$yH$_2$O}
\author{J.-P. Rueff}
\affiliation{Synchrotron SOLEIL, L'Orme des Merisiers, Saint-Aubin, BP~48, 91192 Gif-sur-Yvette Cedex, France.} 
\affiliation{Laboratoire de Chimie Physique--Mati\`ere et Rayonnement (UMR~7614), Universit\'e Pierre et Marie Curie, 11 rue Pierre et Marie Curie, 75231 Paris Cedex~05, France.}
\author{M. Calandra}
\affiliation{Institut de Min\'eralogie et Physique de la Mati\`ere Condensée, CNRS - Universit\'{e} Paris 6, Place Jussieu, 75005 Paris, France.}
\author{M. d'Astuto}
\affiliation{Institut de Min\'eralogie et Physique de la Mati\`ere Condensée, CNRS - Universit\'{e} Paris 6, Place Jussieu, 75005 Paris, France.}
\author{Ph. Leininger}
\affiliation{Laboratoire de Chimie Physique--Mati\`ere et Rayonnement (UMR~7614), Universit\'e Pierre et Marie Curie, 11 rue Pierre et Marie Curie, 75231 Paris Cedex~05, France.}
\author{A. Shukla}
\affiliation{Institut de Min\'eralogie et Physique de la Mati\`ere Condensée, CNRS - Universit\'{e} Paris 6, Place Jussieu, 75005 Paris, France.}
\author{A. Bossak}
\affiliation{European Synchrotron Radiation Facility, BP 42 Grenoble, France.}
\author{M. Krisch}
\affiliation{European Synchrotron Radiation Facility, BP 42 Grenoble, France.}
\author{H. Ishii}
\affiliation{National Synchrotron Radiation Research Center, Hsinchu 30076, Taiwan.}
\author{Y. Cai}
\affiliation{National Synchrotron Radiation Research Center, Hsinchu 30076, Taiwan.}
\author{P. Badica}
\affiliation{Institute for Materials Research, Tohoku University, Katahira 2-1-1, Aoba-ku, Sendai 980-8577, Japan.}
\author{T. Sasaki}
\affiliation{Institute for Materials Research, Tohoku University, Katahira 2-1-1, Aoba-ku, Sendai 980-8577, Japan.}
\author{K. Yamada}
\affiliation{Institute for Materials Research, Tohoku University, Katahira 2-1-1, Aoba-ku, Sendai 980-8577, Japan.}
\author{K. Togano}
\altaffiliation{Present address: National Institute of Materials Science, Tsukuba, 1-2-1 Sengen, 305-0047, Japan.}
\affiliation{Institute for Materials Research, Tohoku University, Katahira 2-1-1, Aoba-ku, Sendai 980-8577, Japan.}

\begin{abstract}
We report on first investigation of the lattice dynamics in the novel superconducting material Na$_{0.35}$CoO$_2$$\cdot$1.3H$_2$O and the non-hydrated parent compound Na$_{0.7}$CoO$_2$ by inelastic x-ray scattering. The measured phonon dispersion along the $\Gamma-M$ direction show a marked softening with hole doping of two optical phonon branches close to the Brillouin zone boundary. The phonon spectra, dispersion, and softening are well reproduced by first-principle calculations. The calculations indicates that the soft branches are mainly composed of Co-vibration modes. The estimation of the critical temperature based on electron-phonon coupling mechanism undisputedly points to a non-conventional superconducting state in this material.  
\end{abstract}
\maketitle

As new superconductors are discovered, the identification of underlying pairing mechanisms is a key step in the understanding of the physics of these materials~\cite{Harlingen1995,Mackenzie2003}. In the hydrated bilayer cobaltate~\cite{Takada2003} Na$_x$CoO$_2$$\cdot$yH$_2$O superconductivity is observed in a narrow range of Na concentration with an optimal $T_c$ of $\sim$5 K~\cite{Chen2004}. The novel material is prepared from the hydration of the parent compound Na$_{0.7}$CoO$_2$ (NCO). The structure~\cite{Takada2003,Takada2004,Lynn2003} of Na$_{0.7}$CoO$_2$ (space group P6$_3$/mmc) consists of alternating layers of Na and CoO$_2$ planes, stacked along the $c$ axis in the hexagonal structure, showing some similarities with high-T$_c$ cuprates. The Na atoms can occupy two crystallographic sites ($2b$ or $2d$), not necessarily in an ordered pattern. In contrast to NCO, Na$_{0.35}$CoO$_2$$\cdot$1.3H$_2$O (NCOH) has a considerably expanded $c$ axis due to the intercalation of ice-like sheets in the structure.

The parent unhydrated compounds Na$_x$CoO$_2$ have a rich phase diagram, ranging from a Curie-Weiss metal ($0.5< x<1$), a charge transfer insulator ($x=0.5$) to a normal paramagnetic metal ($x<0.5$)~\cite{Foo2004}. As it turns out, the Co electronic properties are not fully characterized. In NCO and NCOH, Co supposedly coexists in two formal valent states, Co$^{3+}$ ($S$=0) and Co$^{+4}$ ($S$=1/2), both in low-spin configuration. Thus, the Co ions form a frustrated two-dimensional triangular lattice as opposite to the Cu square lattice in cuprates. The overall picture is further complicated by the highly controversial electronic structure. A crucial issue is the existence of ``hole pockets'' on the Fermi surface : Hole doping results in the depletion of the Co-$t_{2g}$ conduction band. Under trigonal distortion, the $t_{2g}$ triplet splits into two $e_g'$ bands and one $a_g$ band, which cross at the Fermi energy. The latter forms a large hole-type Fermi surface around $\Gamma$, but the main contribution to the hole density of states at the Fermi energy is due to the Co-$e_g'$ holes, supposedly distributed in 6 small pockets. These were predicted by calculations~\cite{Johannes2004a,Singh2000,Zhang2004} and, in the case of non-conventional pairing, their presence could help to distinguish between different pairing symmetries~\cite{Mazin2005}. However they have not be detected in photoemission experiments~\cite{Hasan2004,Yang2004}. More recently, the formation of hole pockets are found robust against dynamical Coulomb correlations~\cite{Ishida2005}, while no such a Fermi-surface topology is reported in the strong coupling limit~\cite{Zhou2005}. 
The proximity of charge, magnetic~\cite{Boothroyd2004} and structural~\cite{Huang2004} instabilities and the frustrated Co-O triangular lattice suggests that the novel cobaltate could have exotic superconducting states.   No information (experimental or theoretical) exists on phonon dispersion and electron-phonon coupling in this material and the question remains of whether a conventional electron-phonon mechanism would yield the measured $T_c$.

In this work, we provide concrete evidence for the exotic character of the superconducting state. We measure the phonon dispersion in NCO and NCOH single crystals by inelastic x-ray scattering (IXS). The phonon dispersion and the electron-phonon coupling are calculated from density functional theory. The results reveal pronounced softening of two optical phonon modes mainly formed by Co vibrations, while the calculated electron-phonon coupling strength definitively excludes conventional pairing mechanism. Further, the softening corresponds closely to what one would expect from a nesting mechanism arising from the $e_{g}^{\prime}$ hole-pocket Fermi surface. These findings go a long way in identifying the character of the superconducting state in NCOH~\cite{Mazin2005}. 

The small sample size ($400\times200\times50\ \mu\mbox{m}^3$) excludes neutron scattering while the focused x-ray beam makes IXS the privileged technique for measuring phonons~\cite{Astuto2002}. The experiment was carried on the ID-28 beamline at the European Synchrotron Radiation Facility (ESRF). Details about the experimental setup has been described else\-where~\cite{Krisch2002}. The resolution of 3 meV was achieved by selecting the Si(9,9,9) order of reflection for both the monochromator and the 5 spherical analyzers, and owing to the extreme backscattering geometry of the IXS spectrometer. The monochromatic beam is focused by a toroidal mirror into a spot of $250\times90\ \mu\mbox{m}^2$ (horizontal$\times$vertical) full width at half maximum at the sample position. NCO was mounted in a vacuum chamber at room temperature in order to reduced background scattering, and remove humidity from the air. To preserve hydration, NCOH was mounted in a closed-cycle cryostat and maintained at 100 K during the measurement. The lattice parameters for both samples were estimated by monitoring the (1,0,0) and (1,0,1) Bragg reflections on the detector. We obtained $a$ = 2.82 \AA\ and $c$ = 11.63 \AA\ in NCO and $a$ = 2.83 \AA\ and $c$ = 19.6 \AA\ in NCOH, in good agreement with the values reported in the literature for the two phases~\cite{Lynn2003}. In the latter, we measured a critical temperature $T_c$ of 4.6 K by magnetic susceptibility. All the phonon spectra were measured in the second Brillouin zone around (1,0,0). To extract the phonon frequencies, the experimental spectra were least-square fitted to a sum of Lorenzian pairs, weighted by the Bose factor. 

\begin{figure}\label{fig1}
\includegraphics[width=8.5 cm]{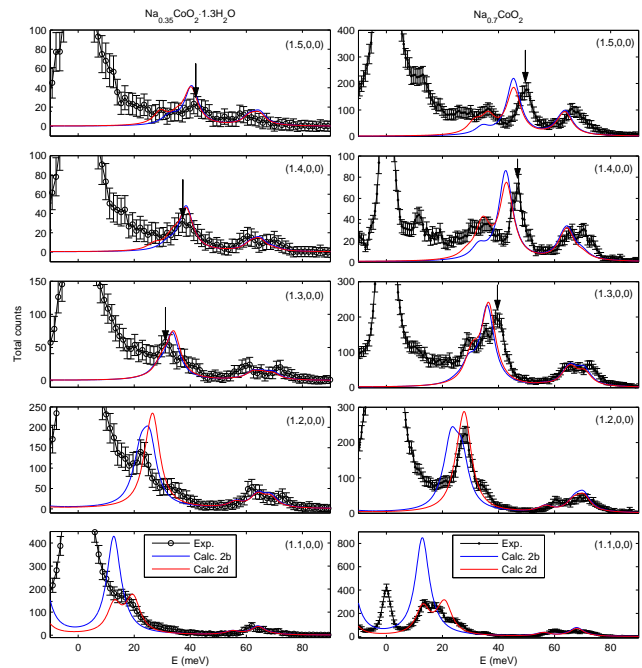}
\caption{IXS spectra in non-hydrated and hydrated compounds measured along the $\Gamma$-$M$ direction (open circles). The measurements are compared to ab-initio calculations of the dynamical structure factor (solid lines). The calculated spectra are shown for Na occupying sites $2b$ and $2d$. The main dispersive feature is pointed by arrows.}
\end{figure} 

Figure~1 shows the IXS phonon spectra in NCOH and NCO, along the $\Gamma$-$M$ direction. Well defined phonon-like excitations are observed up to 80 meV, while non-dispersive features show up in most of the spectra up to the 10 meV region. As discussed hereafter, these are not correlated to any calculated phonon modes of NCOH and NCO, and may be reasonably attributed to the presence of impurities or to Na disorder in the sample. The spectra are compared to first-principle calculations. Electronic structure calculations~\cite{Baroni2001} are performed using Density Functional Theory (DFT) in the local density approximation (LDA) and ultrasoft pseudopotentials~\cite{Vanderbilt1990}. The electronic wavefunctions and the charge density are expanded using a 35 and a 350 Ryd cutoff respectively. The dynamical matrices and the electron-phonon coupling are calculated using Density Functional Perturbation Theory in the linear response~\cite{Baroni2001}. For the electronic integration in the phonon calculation we use a $N_{q}=6\times6\times2$ uniform k-point mesh and and Hermite-Gaussian smearing of 0.05 Ryd. For the calculation of the electron-phonon coupling we use a finer $N_k=25\times 25\times 25$ Monkhorst Pack mesh. The electron-phonon coupling for a phonon mode $\nu$ with momentum $q$ is derived from:
\begin{equation}
\lambda_{{\bf q}\nu} = \frac{4}{\omega_{{\bf q}\nu}N(0) N_{k}} \sum_{{\bf k},n,m} 
|g_{{\bf k}n,{\bf k+q}m}^{\nu}|^2 \delta(\epsilon_{{\bf k}n}) \delta(\epsilon_{{\bf k+q}m})
\end{equation}
where the sum is carried over the Brillouin Zone. $N(0)$ is the electron density 
of states at the Fermi level and $\epsilon_{\bf kn}$ are the DFT energy bands.  
The electron-phonon 
matrix element is $g_{{\bf k}n,{\bf k+q}m}^{\nu}=\langle {\bf k}n|\delta V/\delta u_{{\bf q}\nu} |{\bf k+q} m\rangle /\sqrt{2 \omega_{{\bf q}\nu}}$, where $u_{{\bf q}\nu}$ and $\omega{{\bf q}\nu}$ are the amplitude of the displacement and the frequency of 
the phonon $\nu$ of wavevector ${\bf q}$ respectively while 
$V$ is the Kohn-Sham potential.
The average electron-phonon coupling then reads as $\lambda=\frac{1}{N_q}\sum_{{\bf q},\nu}\lambda_{{\bf q}\nu}$.   
The doping of NaCoO$_2$ is modeled simulating a charged cell with overall number of electrons corresponding to Na$_{x}$CoO$_2$. Since the positions of Na atoms are not determined we perform different simulations for Na occupying the $2b$ or the $2d$ positions. Calculating the phonon dispersion of Na$_{0.35}$CoO$_2$$\cdot$yH$_2$O is intractable due to the large number of atoms in the cell. However it has been shown~\cite{Johannes2004} that the main effect of the intercalation of water in Na$_{x}$CoO$_2$ is the $c-$axis expansion. Unfortunately calculations of Na$_{x}$CoO$_2$ with an expanded lattice parameters produces very flat bands with two very steep unphysical Na bands crossing the Fermi level~\cite{Johannes2004}. As a consequence, the phonons were calcualted in Na$_{0.35}$CoO$_2$ and are here compared with the experiment in Na$_{0.35}$CoO$_2$$\cdot$yH$_2$O. 

Despite the approximations involved in the calculations, there is broad agreement with experiment as observed in Fig.~1. The measured phonon linewidths are systematically found larger (5-6 meV) than the experimental resolution (3 meV). The broadening effect applies to all modes, independent of the momentum transfer, while calculated phonon linewidths are resolution limited. This points to an extrinsic origin presumably related to the crystal mosaicity ($\sim$ 1$^\circ$), besides other cumulative effects. For comparison the calculated spectra were convoluted with a 6 meV broadening. 
 
We focus more particularly on the optical phonon mode (indicated by arrows in Fig.\ 1) which is dispersive along the $\Gamma$-$M$ direction. The theoretical calculations correctly reproduce the frequency and spectral weight of this mode, although the calculated frequencies in NCO are slightly lower than measured ones close to $M$. This discrepancy can be easily understood when taking into account uncertainty in the determination of the Na concentration ($x=0.7\pm 0.1$) and the strong sensitivity of the calculated phonon frequency of this mode on doping, in the vicinity of the $M$ point. Using $x=0.8$ for instance, we obtain $\omega=47.6$ meV at $M$, within 5\% of the experiment. The same mode is seen in the hydrated sample (see arrows), but now significantly softened. At the $M$ point, the experimental softening amounts to $\Delta\omega=7.6\pm1$ meV.

The experimental phonon frequencies are reported in Fig.~2 for NCO and NCOH along with the full calculated dispersion in model Na$_{0.7}$CoO$_2$ and Na$_{0.35}$CoO$_2$ compounds. The phonon dispersion is obtained by Fourier interpolation of the dynamical matrices computed on the $N_q$ points mesh. The intensity of the dynamical structure in the measured Brillouin zone is gray scale-coded. Raman and infrared data~\cite{Li2004a} are also indicated as reference. The theoretical phonon dispersions resemble each other closely except for the optical branches in the 40--45 meV energy region which are sensibly softened on reducing $x$ from 0.7 to 0.35. The other branches are barely affected. The softening is consistently reproduced by the experimental results in the hydrated sample despite the mediocre quality of the spectra. Our calculation indicates that the softening is due principally to the reduction of Na content and not to hydration. In addition to the calculations in Na$_{0.35}$CoO$_2$ with a reduced $c$ parameter (shown in Fig.\ 2a), the effect of hydration was modeled by simulating two layers of CoO$_2$ separated by an expanded $c$ parameter and having a formal charge corresponding to $x=0.35$. The calculated phonon dispersion (not shown here) exhibit a comparable softening of the optical branches, which signifies that softening is solely related to the doping in the CoO$_2$ layers. 
As inferred from the eigenvector analysis of the computed dynamical matrix, in the two compounds the high energy region of the phonon dispersion is mainly composed of oxygen vibrations while the intermediate energy region which shows the main softening effect is mainly due to cobalt vibrations.

\begin{figure}
\label{fig2}
\includegraphics[width=8.5 cm]{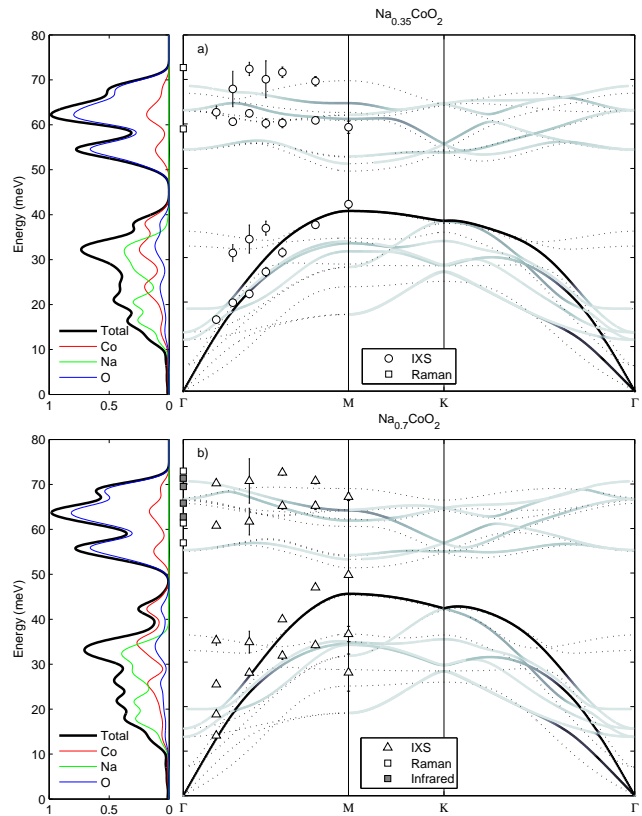}
\caption{Measured (open symbols) and calculated (solid circles and dotted lines) phonon dispersion along high-symmetry directions in NCOH and NCO. For clarity purpose, only calculations with Na-$2b$ sites occupied are shown. The structure factor intensity is indicated in normalized unit by the gray scale form black (1) to light gray (10$^{-6}$), and below 10$^{-6}$ as dotted lines. Raman and infrared data are borrowed from Ref.~\onlinecite{Li2004a}. Left panels represent the projected phonon density of states.}
\end{figure}

We can now estimate the critical temperature for Na$_{0.7}$CoO$_2$ and for Na$_{0.35}$CoO$_2$ from the average electron-phonon coupling $\lambda$ and phonon frequency logarithmic average $\langle\omega\rangle$. The calculated values are $\lambda=0.21$ and  $\langle\omega\rangle=42.5$ meV for $x=0.7$ and $\lambda=0.29$ and $\langle\omega\rangle=40.3$ meV for $x=0.35$. Using McMillan's formula we obtain $T_c\approx 0.001$ K for $x=0.7$ and $T_c=0.15$~K for $x=0.35$ with the Coulomb pseudo-potential constant set at $\mu^{*}=0.1$. These are upper estimates of $T_c$ since in both systems $\mu^*$ is probably larger. Calculations for the expanded CoO$_2$ bilayer lead to $\lambda=0.28$, $\langle\omega\rangle=38.6$ meV and $T_c=0.12$~K, showing that the expansion of the lattice spacing weakly affects the electron-phonon coupling. Some earlier works indicate that water intercalation affects the valence of Co~\cite{Milne2004,Takada2004}. The agreement between the phonon spectra and our calculation indicates that this effect if any is negligible. Since the calculated $T_c$, based on a conventional electron-phonon mechanism, gives results which are more than an order of magnitude smaller than the measured $T_c$, we have a strong argument for invoking {\it a non-conventional pairing mechanism}.
As noted in Ref.~\onlinecite{Mazin2005}, the determination of the pairing symmetry also requires a knowledge of the electronic structure. In particular a crucial aspect is the presence of hole pockets on the Fermi surface. As it turns out, our phonon spectra reveal information about these hole pockets through their influence on phonon frequencies. Indeed the softening of the optical phonon-mode is connected to the real part of the phonon self-energy due to the electron-phonon interaction, namely:

\begin{equation}
\Pi_{\nu}({\bf q},\omega_{{\bf q}\nu})=2\sum_{{\bf k},m,n} 
|g_{{\bf k}n,{\bf k+q}m}^{\nu}|^2\,
\frac{f_{{\bf k}+{\bf q}m} - f_{{\bf k}n}}
{\epsilon_{{\bf k}+{\bf q}m}-\epsilon_{{\bf k}n}-\omega_{{\bf q}\nu}-i\eta}
\end{equation}

where $f_{{\bf k}m}$ stands for the Fermi function. 
If the matrix elements $g_{{\bf k}n,{\bf k+q}m}^{\nu}$ are assumed
to be constant, $\Pi_{\nu}({\bf q},\omega_{{\bf q}\nu})$ is 
related the one-electron susceptibility as calculated in Ref.~\onlinecite{Johannes2004a}.
For $x=0.35$ the real part of $\Pi_{\nu}({\bf q},\omega_{{\bf q}\nu})$ displays pronounced features at the $M$ point which is close to the hole pocket nesting vectors. The net effect on the phonon spectrum is the softening of the phonon branches mentioned earlier.

In conclusion, we present first measurements and calculations of the phonon dispersion of Na$_{0.7}$CoO$_2$ and Na$_{0.35}$CoO$_2$$\cdot$yH$_2$O using inelastic scattering of hard x-rays as a probe of bulk properties. A good agreement is found between theoretical and experimental phonon dispersion. The estimation of the superconducting critical temperature assuming a phonon-mediated mechanism leads to values which are more than an order of magnitude smaller than the measured critical temperature. Moreover the measured and calculated softening of optical phonon modes can be traced back to the existence of the hole pocket Fermi surface.  These results point in the direction of a non-conventional superconducting state in Na$_{0.35}$CoO$_2$$\cdot$yH$_2$O.


\end{document}